\documentclass{ws-mpla}
\usepackage{amsmath, amssymb, graphics, graphicx, hyperref, threeparttable}
\usepackage{natbib}
\bibpunct{(}{)}{;}{a}{}{,}

\begin{document}

\title{On the detection of point sources in Planck LFI 70 GHz CMB maps based on cleaned K-map}

\author{H.G. Khachatryan$^{1,2,3}$ \and A.L. Kashin$^{2,3}$ \and E. Poghosian$^{2,3}$ \and G. Yegorian$^{2,3}$}
\address{$^{1}$Laboratoire d'astrophysique, Ecole Polytechnique F\'ed\'erale de Lausanne (EPFL), Observatoire de Sauverny, CH 1290 Versoix}
\address{$^{2}$Center for Cosmology and Astrophysics, Alikhanyan National Science Laboratory, Alikhanyan brothers 2, Yerevan, Armenia}
\address{$^{3}$Yerevan State University, A. Manoogian 1, Yerevan, Armenia}
\maketitle

\begin{abstract}
We use the Planck LFI 70GHz data to further probe point source detection technique in the sky maps of the cosmic microwave background 
	(CMB) radiation. The method developed by Tegmark et al.\ for foreground reduced maps and the Kolmogorov parameter as the descriptor are adopted for the analysis of Planck satellite CMB temperature data. Most of the detected points coincide with point sources already revealed by other methods. However, we have also found 9 source candidates for which still no counterparts are known.
\end{abstract}

\section{Introduction}

New data by Planck experiment provide more accurate information on the cosmic microwave background (CMB) \citep{PlanckI2013,Bouchet14}. The latter is known to carry crucial information on the properties in the early Universe, while the temperature fluctuations in the CMB maps are basically Gaussian \citep{Komatsu2003,Ferg2012,PlanckXXIII}. Along with that, certain non-Gaussian properties as the Cold Spot, the low multipole anomaly, etc \citep{Cruz2005,Vielva2010,CMB-anomality} have been also noticed, including in the Planck data \citep{VG2014,Rassat2014}. Among the important issues regarding CMB, is the separation of point sources (galaxies, quasars, blazars, GRBs, etc) in the data. Here we study this issue in the Planck data using the method of Kolmogorov stochasticity parameter. 
    
The cosmic microwave background (CMB) maps have been examined for the detection of point sources as potential foregrounds, which could be either cosmological or Milky Way objects emitting thermally or non-thermally \citep{PlanckI2013,Planck13}. A corresponding catalog is made available in \href{http://irsa.ipac.caltech.edu/Missions/planck.html}{NASA/IPAC Infrared Science Archive}.  Various methods, including the wavelets and needlets, have been used for separation of point sources in pixelized sky maps (see, e.g., \citet{Curto13,Sc,Bat,Lanz2013}).  Most of the detected sources in CMB maps coincide with known radio sources, quasars, blazars, although some sources remain still unidentified \citep{Wright,Jarosik2011}.

The Kolmogorov stochasticity parameter (KSP) has been already involved for such aims \citep{GK2008,G2010} using Wilkinson Microwave Anisotropy Probe (WMAP) data: some of the sources revealed by that method initially had no counterparts, however their counterparts have been detected later by the Fermi satellite as gamma sources and were included in Fermi LAT 1-year Point Source Catalog \citep{Fermi1}. In the present study we continue the application of the KSP technique aiming to detect point sources in Planck LFI 70GHz data set \citep{LFI2011}, and for that goal we borrow the maps cleaned by the method developed in \cite{Tegmark03}.
 
This paper is organized as follows. First we introduce briefly the Kolmogorov method, then we construct cleaned K-maps using the Tegmark et al.\ method, modified for such maps.  Then we use the resulting maps to reveal the point sources and other non-Gaussian structures on the CMB map.

\section{Kolmogorov method and CMB maps}

In his original work Kolmogorov \citet{Kolm} has introduced a method for determining whether a given random number sequence $X_i,\,{i=1,..,N}$, ordered in an increasing way, obeys a given statistic or not \citep{Arnold, Arnold_ICTP, Arnold_UMN, Arnold_MMS,Arnold_FA}. For such a purpose a so-called empirical cumulative distribution function $F_N(x)$ is calculated,
\begin{eqnarray}
  F_N(x)=
  \begin{cases}
    0\ , & x < X_1\ \\
    k/N\ , & X_i \le x,\ \ k = 1,2,\ldots,N-1 \\
    1\ , & X_N \le x\,
  \end{cases}
\end{eqnarray}
where $k$ is the number of elements which obey to the relation $X_i\le x$. Having assumed a particular theoretical cumulative
distribution function (CDF) $F(x)$, the parameter $\lambda_N$ is easily calculated,
\begin{equation}
  \lambda_N = \sqrt{N}\, \sup_{x} \left| F_N(x)-F(x) \right|.
\end{equation}
Kolmogorov proved that in the limit $N\rightarrow\infty$, $\lambda_N$, which is a random variable, has a cumulative distribution
function $\Phi(\lambda)$ reading as
\begin{equation}
  \Phi(\lambda)=\sum_{j=-\infty}^{+\infty}{(-1)^{j}e^{-2j^{2}\lambda^{2}}}.
\label{distK}
\end{equation}
The function $\Phi(\lambda)$ can be expressed as a particular value of theta functions, since
$\Phi(\lambda)=\vartheta_{4}(0,e^{-2{\lambda}^2})=\vartheta_{3}(0,-e^{-2{\lambda}^2})$ \citep[see][16.27]{Abramovitz}.

In the series of papers \citep{Arnold, Arnold_ICTP, Arnold_UMN, Arnold_MMS,Arnold_FA} Arnold examines the application of this statistic to determine degree of randomness of a given sequence. Although this statistic was intended for i.i.d sequences, he applied this statistic to arithmetical and geometrical progressions and revealed the relative degree of randomness e.g. of arithmetical progression with respect to  geometrical one.

\begin{figure}[t]
\includegraphics[width=0.95\textwidth]{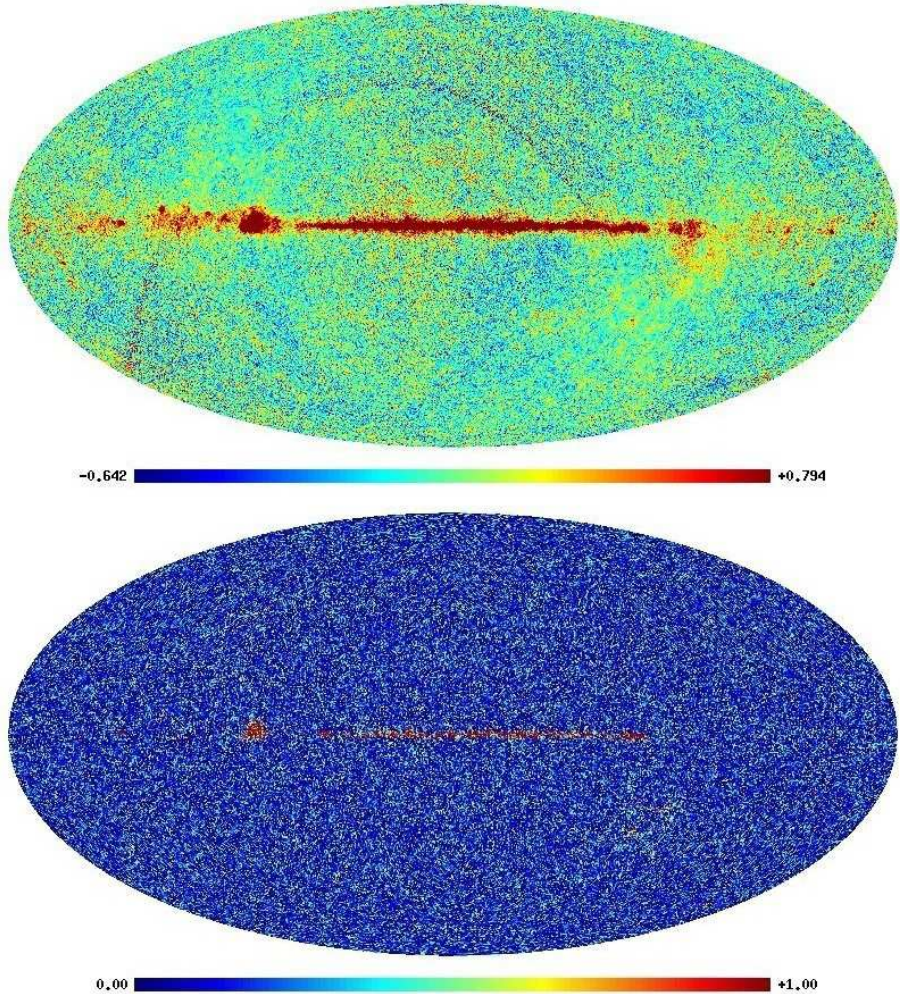}
\caption{Planck LFI 70GHz Cosmic microwave background temperature map with resolution parameter $n_{\rm side}=2048$ (top) and K-maps with resolution parameters $n_{\rm side}=128$ (bottom). It can be seen a round spot on the CMB temperature map affected by inhomogeneous coverage of the sky. This spot appears on K-map as a region with unusual high value of KSP (see Fig.~\ref{cKSP} and Figs.~10,15,18 of \citet{LFI2011}).}
\label{Maps}
\end{figure}

\section{Cleaned CMB K-map}

\subsection{Planck LFI 70GHz CMB and Kolmogorov maps}
It is well known that the CMB sky has basically Gaussian distribution, although along with certain non-Gaussian features (see e.g. \citet{Bernui2010,Planck13a,Spergel2000}). Based on this empirical fact, the Gaussian distribution is used in Kolmogorov method to construct a K-map. In the case of Planck LFI 70 GHz CMB map \citep{LFI2011} with resolution parameter $n_{\rm side}=2048$, for every compact region of the CMB map containing 256 neighboring temperature pixels, one corresponding pixel value of the K-map is obtained. But for Planck LFI 70 GHz CMB map with resolution parameter $n_{\rm side}=1024$, the same procedure is done for 64 neighboring temperature pixels. Therefore for both CMB temperature maps with resolution parameters $n_{\rm side}=2048,1024$ we get K-maps with resolution parameter $n_{\rm side}=128$. This adds some statistical inhomogeneity in our calculations but provides more maps for Tegmark et al. foreground reducing method. For both Planck LFI 70 GHz CMB maps with $n_{\rm side}=2048$ and $n_{\rm side}=1024$ resolution parameters (details in \citet{Healpix}), we obtain a K-map with $n_{\rm side}=128$ parameter. This means that if the CMB map has $n_{\rm map}=12n_{\rm side}^2=50331648$ pixels, then K-map has $n_{\rm map}=196608$ pixels. This is due to the fact that KSP is a statistical parameter. Then the KSP distribution maps over the whole sky can be obtained for Planck LFI 70 GHz data. It is seen in Fig.~\ref{Maps} that the Galactic disk region has higher and saturated KSP values, which indicate that it has a non-Gaussian distribution, as distinct from the CMB. Also a lot of pixels have a high value of KSP, but most of them, as it will be shown below, are due to instrumental and other types of noise, also of non-Gaussian nature.

\begin{figure}[t]
\includegraphics[width=0.95\textwidth]{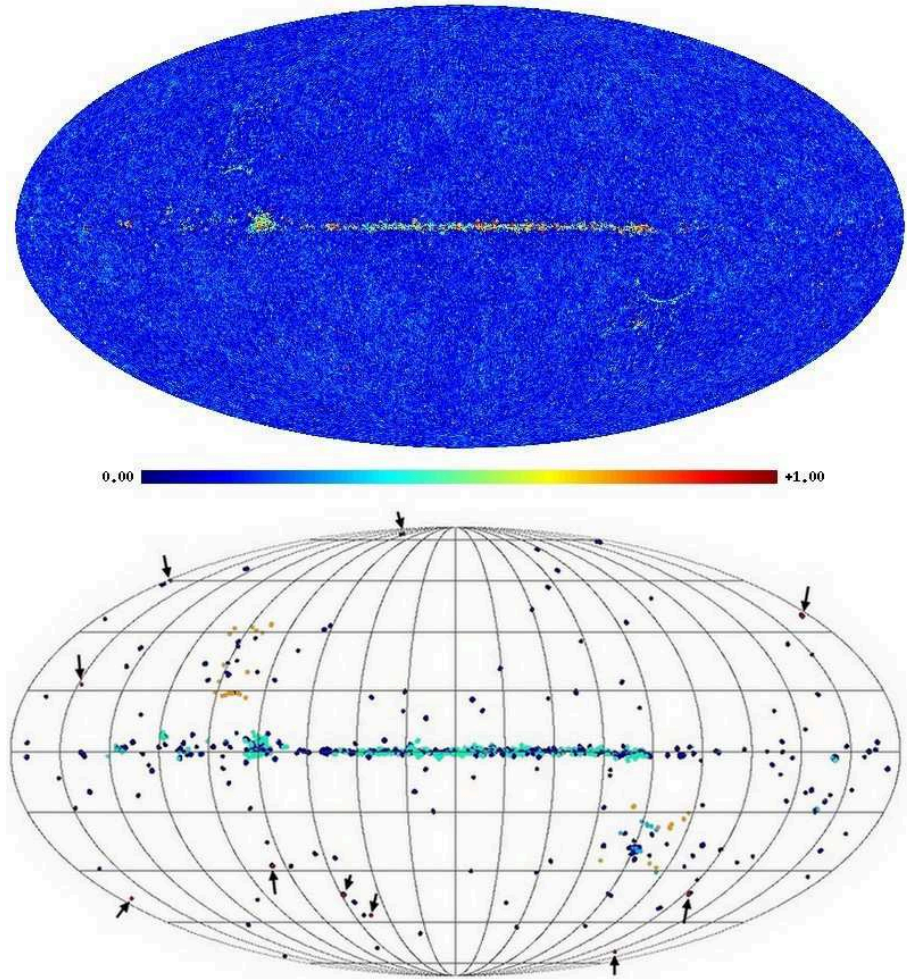}
\caption{Planck LFI 70GHz foreground reduced K-map \citet{Tegmark03} with resolution parameters $n_{\rm side}=128$ (top) and regions with higher value of KSP ($\Phi>0.5$). A subtle star-like shape of inhomogeneous sky coverage pixels can be seen in both foreground reduced K-map and among higher KSP values of KSP indicated as yellow dots (see Fig.~5 of \citet{PlanckI2013}). Blue dots of high KSP values indicate those non-Gaussian regions that have counterpart in Planck Compact Point Source Catalog. Red (pointed by arrows) and yellow dots are those having no counterpart and regions affected by inhomogeneous sky coverage, respectively.}
\label{cKSP}
\end{figure}

Throughout the analysis we mainly use the HEALPIX program package for calculation of the spherical harmonic coefficients and to reconstruct the maps \citet{Healpix}. For some tedious manipulating procedures with the spherical harmonic coefficients $a_{lm}$ (see Eq.~(\ref{harm}) below) we use the GLESP program package \citet{GLESP}. For maps with low resolution (in our case for K-map, $n_{\rm side}=128,n_{\rm map}=196608$) the $a_{lm}$ calculation and map reconstruction the use of HEALPIX program package led to accuracy problems. To test whether the calculation error is small or not for a low resolution map, we construct a unity HEALPIX map with $n_{\rm side}=128$, and another one with $n_{\rm side}=2048$, and run them through the $a_{lm}$ calculation and the map reconstruction procedures. For both cases this procedure adds some non-isotropic noise which has almost zero mean ($e$) and a very small standard deviation ($\sigma$). We obtain $\langle e_{2048}\rangle= 10^{-7}$, $\left\langle e_{128}\right\rangle= 10^{-4}$, $\sigma_{2048}=5.6\cdot 10^{-5}$, $\sigma_{128}=9\cdot 10^{-4}$. Although some pixels around the poles have bigger error, the fraction of these pixels on the sky for $n_{\rm side}=128$ is less than 0.5\%. So for $n_{\rm side}=128$, the error and standard deviation are sufficiently small, which enables one to calculate the $a_{lm}$ and to construct the cleaned map for KSP using Tegmark et al.\ method.

\subsection{Modified Tegmark et al. method}
Any full sky map can be represented via a series of Legendre spherical functions $Y_{lm}(\theta,\varphi)$ \citep{Hivon2002}, where 
\begin{eqnarray}
T(\theta ,\varphi )&=&\sum_{l,m}a_{lm}Y_{lm}(\theta ,\varphi ), \nonumber \\
a_{lm}&=&\int T(\theta ,\varphi ) \, Y^{\ast}_{lm}(\theta ,\varphi )\sin \theta \, d\theta \, d\varphi .
\label{harm}
\end{eqnarray}
As easily shown, the coefficients $C_l$ of the Legendre polynomials $P_l(\cos\theta)$ in the two point correlation function $C(\theta)$ of the power spectrum are related to the $a_{lm}$ by
\begin{eqnarray} 
  C_{l}&=& \langle a_{lm}^{\ast}a_{lm}\rangle , \nonumber \\
  C(\theta )&=&\frac{1}{4\pi }\sum_{l,m}\left(2l+1\right) C_{l}\,P_{l}(\cos \theta ).
\label{Clr}    
\end{eqnarray}
So called cross power spectra is a common technique described in \citet{Hinshaw03}. It can calculated via taking the cross-correlation power spectrum coefficients for different type of $a_{lm}$, as
\begin{equation}
\tilde{C}_l^{ij}=\langle a_{lm}^{{\ast}i}a_{lm}^{j}\rangle.
\end{equation}
Tegmark et al. \citep{Tegmark03} CMB foreground reducing method uses cross-correlation power spectrum matrix to obtain weights for every band. This method is commonly used for obtaining a CMB foreground-reduced map from the original WMAP and Planck (SMICA composite map) CMB maps. In this method every map from different bands is weighted with weights $w_{l}^{i}$. Here $l$ is the multipole number and $i$ the band index. It differs from the interlinear combination (ILC) method of weighting different bands suggested by the WMAP team \citep{Jarosik2011} by dependence of weights on the multipole numbers.

We use Tegmark et al.\ method \citep{Tegmark03,Saha06,Saha08,VG2009b} to develop a cleaned CMB K-map from three couples of Planck LFI 70 GHz maps: nominal, nominal\_ringhalf1, nominal\_ringhalf2. Each element in couple of maps contains both maps with resolution parameter $n_{\rm side}=1024,2048$. This method, based on the power spectrum comparison, assigns a weight $w_{l}^{i}$ for each map $i$ and multipole $l$. In order not to distort the original map, $w_{l}^{i}$ should obey the relation
\begin{equation}
\sum_{i}{w_{l}^{i}}=1. 
\label{restriction}
\end{equation}
But in our case it is mandatory to preserve original K-maps mean value because Kolmogorov parameter varies from zero to unity. Because first harmonic coefficient $a_{00}$ depends only on mean value of map, we take $a_{00}$ of cleaned map as a mean value over original maps $a_{00}$ values
\begin{equation}
\acute{a_{00}}=\left\langle{a_{00}}\right\rangle.
\label{restriction0}
\end{equation}
\begin{figure}[t]
\includegraphics[width=0.95\textwidth]{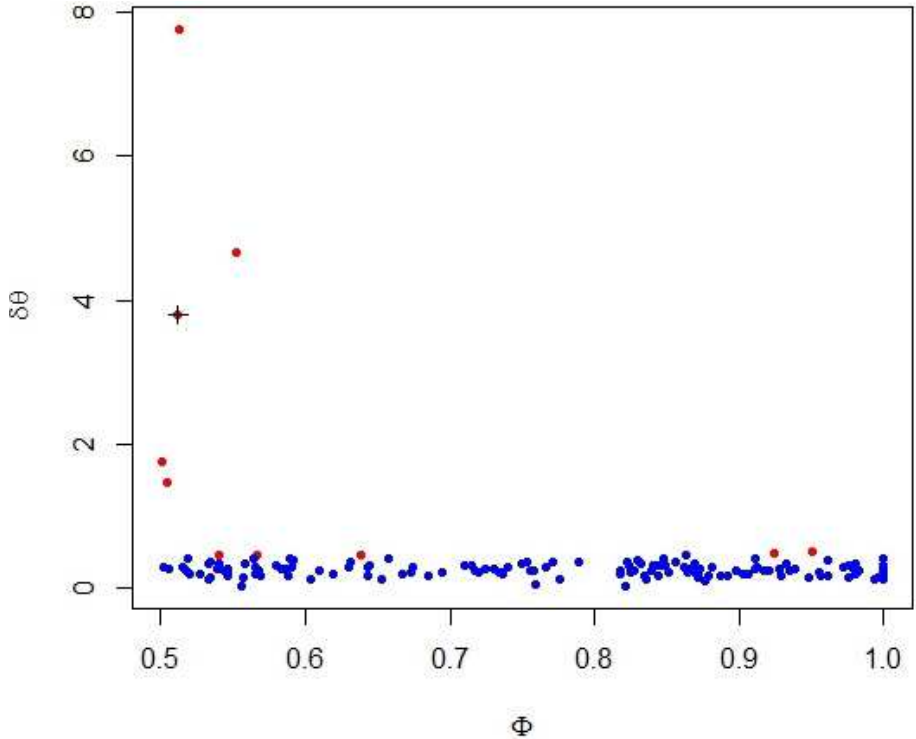}
\caption{Angular distance between the nearest point source of Planck LFI 70 GHz compact point source catalog and KSP high value pixel ($d\theta$) vs. Kolmogorov  parameter $\Phi$ outside the Galactic Region $|b|<20$. Red and blue dots are KSP high value pixels whose distance smaller and larger than $0.45$, one pixel angular diameter of HEALPIX map with resolution parameter $n_{\rm side}=128$, respectively. Black cross ($+$) over a red dot indicate K-map pixel around which too few observations are available. This pixel located over the spot is mentioned in Fig.~\ref{Maps}.}
\label{points}
\end{figure}
In the Kolmogorov method, if a sequence of random numbers $T_n$ obeys a theoretical distribution function $F(x)$ then $N$ realizations of this sequence gives $\Phi_{N}$, $\lambda_{N}$. The remarkable point of the method is that, $\Phi_{N}$ has a uniform distribution and the mean value $\left\langle \Phi\right\rangle=0.5$. Therefore for a CMB K-map the value $0.5$ is a natural threshold for distinguishing non-Gaussian areas from others, since Gaussian temperature pixels cannot have $\Phi> 0.5$. Further, the cleaned K-map mean value and sigma are respectively $\left\langle \Phi\right\rangle=0.17$, $\sigma_{\Phi}=0.09$, so pixels with $\Phi>0.5$ mostly exceed the 3-$\sigma$ region. 
\section{Non-Gaussian areas and point sources}

\subsection{Inhomogeneous coverage of the sky and cleaned K-map}
It is well-known that Planck satellite covered the sky inhomogeneously. It can be seen in Fig.~5 of \citet{PlanckI2013} that certain regions on the sky observed about thousand times but there are also sky regions that were not covered (observed) at all. Kolmogorov parameter is a statistical parameter therefore it is sensible to such inhomogeneous coverage, too. So if a compact region of 256 temperature neighboring pixels is located on the border where number of observations (NOBS) decrease or increase sharply, then we have high value of KSP. These NOBS affected regions are colored yellow to indicate them among other regions with high value of KSP ($\Phi>0.5$) in Fig.~\ref{cKSP}. In contrary with these yellow colored pixels where number of observations decreases form very high values to mean value, there is another pixel of K-map around which hardly any observation was done (coordinates on the K-map $l=173.632$, $b=-50.48$, colored red). So Kolmogorov parameter detects inhomogeneous coverage of the sky,too.    

\subsection{Galactic region and point sources}
In Fig.~\ref{cKSP} one can see the Galactic disk, the Large Magellanic Cloud (LMC) and other possible point sources. There are in total 1321 non-Gaussian pixels among 196608 (about $0.7\%$). Only 153 among them are outside the Galactic disk region ($|b|<20$), the Large Magellanic Cloud ($l=280.4136$, $b=32.9310$) and regions of sky affected by inhomogeneous sky coverage (so called "NOBS"). So, these 153 regions are point source candidates. Indeed, most of them are found in the catalog of \citet{Planck-sc} (see Fig.~\ref{points} for more details). Note that even if one takes Galactic region as $|b|<18$, then it would just add more 3 point sources to these 153. Hence these 153 point source candidates depend very slightly on Galactic disk mask choice. Among that 153 point source candidates only nine possible point sources remain unidentified in any of the known catalogs. 

We are comparing and confirming that our possible point source candidates have counterparts in Planck Compact Source Catalog in case when angular distance between a point source is equal or less than the average angular size of HEALPIX pixel. For a map of resolution parameter $n_{\rm side}=128$ the HEALPIX pixels angular size is $d_{pix}=\acute{27.5}$. But this restriction is very strict and some cases can be when point sources have large angular size. Therefore five of these unidentified point sources being close neighbors ($\delta\theta<1$) might be also affected by Planck point sources. Remaining four unidentified point source candidates have counterparts in Two Micron All Sky Survey (2MASS) galaxy catalog. 

\begin{threeparttable}
\begin{center}
\caption{Coordinates of possible point sources detected in cleaned K-map.}
\label{ps}
\begin{tabular}{ccccc}
\footnotesize{l} & \footnotesize{b} & \footnotesize{KSP value} & \footnotesize{$\delta\theta$} & \footnotesize{2MASS H-mag}\\ \hline 
\footnotesize{87.3529} & \footnotesize{83.7837} & \footnotesize{0.64} & \footnotesize{0.47} & \footnotesize{--$^b$}\\ 
\footnotesize{158.555} & \footnotesize{22.0243} & \footnotesize{0.51} & \footnotesize{7.76} & \footnotesize{$13.174^a$}\\ 
\footnotesize{178.333} & \footnotesize{60.0563} & \footnotesize{0.50} & \footnotesize{1.76} & \footnotesize{$16.144^a$}\\ 
\footnotesize{184.195} & \footnotesize{45.784} & \footnotesize{0.92} & \footnotesize{0.49} & \footnotesize{--$^b$}\\  
\footnotesize{50.0562} & \footnotesize{-57.0185} & \footnotesize{0.504} & \footnotesize{1.47} & \footnotesize{$14.676^a$}\\ 
\footnotesize{86.1328} & \footnotesize{-38.6822} & \footnotesize{0.95} & \footnotesize{0.51} & \footnotesize{--$^b$}\\ 
\footnotesize{59.0367} & \footnotesize{-49.3128} & \footnotesize{0.57} & \footnotesize{0.47} & \footnotesize{--$^b$}\\  
\footnotesize{217} & \footnotesize{-73.4963} & \footnotesize{0.55} & \footnotesize{4.66} & \footnotesize{$12.207^a$}\\ 
\footnotesize{237.682} & \footnotesize{-48.9228} & \footnotesize{0.54} & \footnotesize{0.47} & \footnotesize{--$^b$}\\ 
\end{tabular}
\begin{tablenotes}
\item[a] \scriptsize{nearby galaxy exists in 2MASS catalog within radius of one arc minute}
\item[b] \scriptsize{unidentified point source but very near to some object in Planck Point Sources catalog \citet{Planck-sc}}
\end{tablenotes}
\end{center}
\end{threeparttable}

\section{Conclusions}

We use the Kolmogorov stochasticity parameter to probe non-Gaussianity of Planck LFI 70 GHz CMB maps. It is shown that for about 72\% of the cleaned map, the Kolmogorov function has values within the interval $0.0<\Phi<0.2$, which implies that the CMB maps are indeed Gaussian with high precision. 

Non-Gaussian pixels (with $\Phi>0.5$) are rather rare ($1321$), i.e., less than 1\%. Although this result is true for cleaned K-map, it shows that different sources add certain non-Gaussian noise into CMB map and they can be removed by foreground reducing methods.

Another remarkable feature of both cleaned and original K-maps is that it shows some areas of inhomogeneous coverage of the sky by Planck satellite. It is obvious that some pixels in K-maps have high value of KSP due to this effect. 

For searching counterparts between these non-Gaussian regions and any available catalogs, we utilize the main feature of HEALPIX map- equal area pixels. Therefore 1-pixel angular size is about $0.45$ degrees in case of resolution parameter $n_{\rm side}=128$. It can be seen in Fig.~\ref{points} that this assumption works very well. K-map pixels outside Galactic disk region ($|b|>20$) with KSP value $\Phi>0.5$ indicate possible point sources and other non-Gaussian regions. Indeed 143 of them have counterparts in Planck Compact Point Sources Catalog of LFI 70 GHz. But nine non-Gaussian regions still have no known counterpart. Note that, a non-Gaussianity appears around the region of CMB sky with extremly low number of observations.

\section*{Acknowledgements}
  We are grateful to V.\ Gurzadyan and colleagues in Center for Cosmology and Astrophysics for numerous comments and discussions. HG would like to express his appreciation to G. Meylan, D. Pfenniger and G. Nurbaeva for helpful comments and support during the early stages of this study and to O.\ Verkhodanov for information about GLESP.


\begin{thebibliography}{0}
\expandafter\ifx\csname natexlab\endcsname\relax\def\natexlab#1{#1}\fi

\bibitem[{{Abdo} {et~al.}(2010){Abdo}, {Ackermann}, {Ajello}, {Allafort},
  {Antolini}, {Atwood}, {Axelsson}, {Baldini}, {Ballet}, {Barbiellini}, \&
  et~al.}]{Fermi1}
{Abdo}, A.~A., {Ackermann}, M., {Ajello}, M., {et~al.} 2010, ApJS, 188, 405

\bibitem[{{Abramowitz} \& {Stegun}(1970)}]{Abramovitz}
{Abramowitz}, M. \& {Stegun}, I. 1970, Handbook of Mathematical Functions (New
  York: Dover Publications)

\bibitem[{{Ade} {et~al.}(2013{\natexlab{a}}){Ade}, {Aghanim}, {Alves},
  {Armitage-Caplan}, {Arnaud}, {Ashdown}, {Atrio-Barandela}, {Aumont},
  {Aussel}, \& et~al.}]{PlanckI2013}
{Ade}, P.~A.~R., {Aghanim}, N., {Alves}, M.~I.~R., {et~al.} 2013{\natexlab{a}},
  submitted to A\&A, preprint arXiv:astro-ph/1303.5062

\bibitem[{{Ade} {et~al.}(2013{\natexlab{b}}){Ade}, {Aghanim}, {Arg{\"u}eso},
  {Armitage-Caplan}, {Arnaud}, {Ashdown}, {Atrio-Barandela}, {Aumont},
  {Baccigalupi}, \& et~al.}]{Planck13}
{Ade}, P.~A.~R., {Aghanim}, N., {Arg{\"u}eso}, F., {et~al.} 2013{\natexlab{b}},
  submitted to A\&A, preprint arXiv:astro-ph/1303.5088

\bibitem[{{Ade} {et~al.}(2013{\natexlab{c}}){Ade}, {Aghanim},
  {Armitage-Caplan}, {Arnaud}, {Ashdown}, {Atrio-Barandela}, {Aumont},
  {Baccigalupi}, {Banday}, \& et~al.}]{PlanckXXIII}
{Ade}, P.~A.~R., {Aghanim}, N., {Armitage-Caplan}, C., {et~al.}
  2013{\natexlab{c}}, accepted for publication in A\&A, preprint
  arXiv:astro-ph/1303.5083

\bibitem[{{Ade} {et~al.}(2013{\natexlab{d}}){Ade}, {Aghanim},
  {Armitage-Caplan}, {Arnaud}, {Ashdown}, {Atrio-Barandela}, {Aumont},
  {Baccigalupi}, {Banday}, \& et~al.}]{Planck13a}
{Ade}, P.~A.~R., {Aghanim}, N., {Armitage-Caplan}, C., {et~al.}
  2013{\natexlab{d}}, preprint arXiv:astro-ph/1303.5084

\bibitem[{{Ade} {et~al.}(2011){Ade}, {Aghanim}, {Arnaud}, {Ashdown}, {Aumont},
  {Baccigalupi}, {Balbi}, {Banday}, {Barreiro}, \& et~al.}]{Planck-sc}
{Ade}, P.~A.~R., {Aghanim}, N., {Arnaud}, M., {et~al.} 2011, A\&A, 536, A7

\bibitem[{{Arnold}(2008{\natexlab{a}})}]{Arnold}
{Arnold}, V.~I. 2008{\natexlab{a}}, Nonlinearity, 21, T109

\bibitem[{{Arnold}(2008{\natexlab{b}})}]{Arnold_ICTP}
{Arnold}, V.~I. 2008{\natexlab{b}}, ICTP/2008/001, Trieste

\bibitem[{{Arnold}(2008{\natexlab{c}})}]{Arnold_UMN}
{Arnold}, V.~I. 2008{\natexlab{c}}, Uspekhi Mat. Nauk, 63, 5

\bibitem[{{Arnold}(2009{\natexlab{a}})}]{Arnold_MMS}
{Arnold}, V.~I. 2009{\natexlab{a}}, Trans. Moscow Math. Soc., 70, 31

\bibitem[{{Arnold}(2009{\natexlab{b}})}]{Arnold_FA}
{Arnold}, V.~I. 2009{\natexlab{b}}, Funct. Anal. Other Math., 2, 139

\bibitem[{{Batista} {et~al.}(2011){Batista}, {Kemp}, \& {Daniel}}]{Bat}
{Batista}, R.~A., {Kemp}, E., \& {Daniel}, B. 2011, IJMPE, 20, 61

\bibitem[{{Bernui} \& {Rebou{\c c}as}(2010)}]{Bernui2010}
{Bernui}, A. \& {Rebou{\c c}as}, M.~J. 2010, Phys. Rev. D, 81, 063533

\bibitem[{{Bouchet} \& {on behalf of the Planck
  collaboration}(2014)}]{Bouchet14}
{Bouchet}, F.~R. \& {on behalf of the Planck collaboration}. 2014, submitted to
  A\&A, preprint arXiv:astro-ph/1405.0439

\bibitem[{{Cruz} {et~al.}(2005){Cruz}, {Mart{\'{\i}}nez-Gonz{\'a}lez},
  {Vielva}, \& {Cay{\'o}n}}]{Cruz2005}
{Cruz}, M., {Mart{\'{\i}}nez-Gonz{\'a}lez}, E., {Vielva}, P., \& {Cay{\'o}n},
  L. 2005, MNRAS, 356, 29

\bibitem[{{Curto} {et~al.}(2013){Curto}, {Tucci}, {Gonz{\'a}lez-Nuevo},
  {Toffolatti}, {Mart{\'{\i}}nez-Gonz{\'a}lez}, {Arg{\"u}eso}, {Lapi}, \&
  {L{\'o}pez-Caniego}}]{Curto13}
{Curto}, A., {Tucci}, M., {Gonz{\'a}lez-Nuevo}, J., {et~al.} 2013, MNRAS, 432,
  728

\bibitem[{{de Oliveira-Costa} {et~al.}(2004){de Oliveira-Costa}, {Tegmark},
  {Zaldarriaga}, \& {Hamilton}}]{CMB-anomality}
{de Oliveira-Costa}, A., {Tegmark}, M., {Zaldarriaga}, M., \& {Hamilton}, A.
  2004, Phys. Rev. D, 69, 063516

\bibitem[{{Doroshkevich} {et~al.}(2005){Doroshkevich}, {Naselsky},
  {Verkhodanov}, {Novikov}, {Turchaninov}, {Novikov}, {Christensen}, \&
  {Chiang}}]{GLESP}
{Doroshkevich}, A.~G., {Naselsky}, P.~D., {Verkhodanov}, O.~V., {et~al.} 2005,
  Int. J. Mod. Phys. D, 14, 275

\bibitem[{{Fergusson} {et~al.}(2012){Fergusson}, {Liguori}, \&
  {Shellard}}]{Ferg2012}
{Fergusson}, J.~R., {Liguori}, M., \& {Shellard}, E.~P.~S. 2012, JCAP, 12, 32

\bibitem[{{G{\'o}rski} {et~al.}(2005){G{\'o}rski}, {Hivon}, {Banday},
  {Wandelt}, {Hansen}, {Reinecke}, \& {Bartelmann}}]{Healpix}
{G{\'o}rski}, K.~M., {Hivon}, E., {Banday}, A.~J., {et~al.} 2005, ApJ, 622, 759

\bibitem[{{Gurzadyan} {et~al.}(2014){Gurzadyan}, {Kashin}, {Khachatryan},
  {Poghosian}, {Sargsyan}, \& {Yegorian}}]{VG2014}
{Gurzadyan}, V.~G., {Kashin}, A.~L., {Khachatryan}, H., {et~al.} 2014, A\&A,
  566, A135

\bibitem[{{Gurzadyan} {et~al.}(2009){Gurzadyan}, {Kashin}, {Khachatryan},
  {Kocharyan}, {Poghosian}, {Vetrugno}, \& {Yegorian}}]{VG2009b}
{Gurzadyan}, V.~G., {Kashin}, A.~L., {Khachatryan}, H.~G., {et~al.} 2009, A\&A,
  506, L37

\bibitem[{{Gurzadyan} {et~al.}(2010){Gurzadyan}, {Kashin}, {Khachatryan},
  {Kocharyan}, {Poghosian}, {Vetrugno}, \& {Yegorian}}]{G2010}
{Gurzadyan}, V.~G., {Kashin}, A.~L., {Khachatryan}, H.~G., {et~al.} 2010,
  Europhys. Lett., 91, 19001

\bibitem[{{Gurzadyan} \& {Kocharyan}(2008)}]{GK2008}
{Gurzadyan}, V.~G. \& {Kocharyan}, A.~A. 2008, A\&A, 492, L33

\bibitem[{{Hinshaw} {et~al.}(2003){Hinshaw}, {Spergel}, {Verde}, {Hill},
  {Meyer}, {Barnes}, {Bennett}, {Halpern}, {Jarosik}, {Kogut}, {Komatsu},
  {Limon}, {Page}, {Tucker}, {Weiland}, {Wollack}, \& {Wright}}]{Hinshaw03}
{Hinshaw}, G., {Spergel}, D.~N., {Verde}, L., {et~al.} 2003, ApJS, 148, 135

\bibitem[{{Hivon} {et~al.}(2002){Hivon}, {G{\'o}rski}, {Netterfield}, {Crill},
  {Prunet}, \& {Hansen}}]{Hivon2002}
{Hivon}, E., {G{\'o}rski}, K.~M., {Netterfield}, C.~B., {et~al.} 2002, ApJ,
  567, 2

\bibitem[{{Jarosik} {et~al.}(2011){Jarosik}, {Bennett}, {Dunkley}, {Gold},
  {Greason}, {Halpern}, {Hill}, {Hinshaw}, {Kogut}, {Komatsu}, {Larson},
  {Limon}, {Meyer}, {Nolta}, {Odegard}, {Page}, {Smith}, {Spergel}, {Tucker},
  {Weiland}, {Wollack}, \& {Wright}}]{Jarosik2011}
{Jarosik}, N., {Bennett}, C.~L., {Dunkley}, J., {et~al.} 2011, ApJS, 192, 14

\bibitem[{{Kolmogorov}(1933)}]{Kolm}
{Kolmogorov}, A.~N. 1933, Giorn.Ist.Ital.Attuari, 4, 83

\bibitem[{{Komatsu} {et~al.}(2003){Komatsu}, {Kogut}, {Nolta}, {Bennett},
  {Halpern}, {Hinshaw}, {Jarosik}, {Limon}, {Meyer}, {Page}, {Spergel},
  {Tucker}, {Verde}, {Wollack}, \& {Wright}}]{Komatsu2003}
{Komatsu}, E., {Kogut}, A., {Nolta}, M.~R., {et~al.} 2003, ApJS, 148, 119

\bibitem[{{Komatsu} \& {Spergel}(2001)}]{Spergel2000}
{Komatsu}, E. \& {Spergel}, D.~N. 2001, Phys. Rev. D, 63, 063002

\bibitem[{{Lanz} {et~al.}(2013){Lanz}, {Herranz}, {L{\'o}pez-Caniego},
  {Gonz{\'a}lez-Nuevo}, {de Zotti}, {Massardi}, \& {Sanz}}]{Lanz2013}
{Lanz}, L.~F., {Herranz}, D., {L{\'o}pez-Caniego}, M., {et~al.} 2013, MNRAS,
  428, 3048

\bibitem[{{Rassat} {et~al.}(2014){Rassat}, {Starck}, {Paykari}, {Sureau}, \&
  {Bobin}}]{Rassat2014}
{Rassat}, A., {Starck}, J.-L., {Paykari}, P., {Sureau}, F., \& {Bobin}, J. 2014

\bibitem[{{Saha} {et~al.}(2006){Saha}, {Jain}, \& {Souradeep}}]{Saha06}
{Saha}, R., {Jain}, P., \& {Souradeep}, T. 2006, ApJ, 645, L89

\bibitem[{{Saha} {et~al.}(2008){Saha}, {Prunet}, {Jain}, \& T.}]{Saha08}
{Saha}, R., {Prunet}, S., {Jain}, P., \& T., S. 2008, Phys. Rev. D, 78, 023003

\bibitem[{{Scodeller} {et~al.}(2012){Scodeller}, {Hansen}, \& {Marinucci}}]{Sc}
{Scodeller}, S., {Hansen}, F.~K., \& {Marinucci}, D. 2012, APJ, 753, 27

\bibitem[{{Tegmark} {et~al.}(2003){Tegmark}, {de Oliveira-Costa}, \&
  {Hamilton}}]{Tegmark03}
{Tegmark}, M., {de Oliveira-Costa}, A., \& {Hamilton}, A. 2003, Phys. Rev. D,
  68, 123523

\bibitem[{{Vielva}(2010)}]{Vielva2010}
{Vielva}, P. 2010, Advances in Astronomy, 2010

\bibitem[{{Wright} {et~al.}(2009){Wright}, {Chen}, {Odegard}, {Bennett},
  {Hill}, {Hinshaw}, {Jarosik}, {Komatsu}, {Nolta}, {Page}, {Spergel},
  {Weiland}, {Wollack}, {Dunkley}, {Gold}, {Halpern}, {Kogut}, {Larson},
  {Limon}, {Meyer}, \& {Tucker}}]{Wright}
{Wright}, E.~L., {Chen}, X., {Odegard}, N., {et~al.} 2009, ApJS, 180, 283

\bibitem[{{Zacchei} {et~al.}(2011){Zacchei}, {Maino}, {Baccigalupi},
  {Bersanelli}, {Bonaldi}, {Bonavera}, {Burigana}, {Butler}, {Cuttaia}, {de
  Zotti}, {Dick}, {Frailis}, {Galeotta}, {Gonz{\'a}lez-Nuevo}, {G{\'o}rski},
  {Gregorio}, {Keih{\"a}nen}, {Keskitalo}, {Knoche}, {Kurki-Suonio},
  {Lawrence}, {Leach}, {Leahy}, {L{\'o}pez-Caniego}, {Mandolesi}, {Maris},
  {Matthai}, {Meinhold}, {Mennella}, {Morgante}, {Morisset}, {Natoli},
  {Pasian}, {Perrotta}, {Polenta}, {Poutanen}, {Reinecke}, {Ricciardi},
  {Rohlfs}, {Sandri}, {Suur-Uski}, {Tauber}, {Tavagnacco}, {Terenzi}, {Tomasi},
  {Valiviita}, {Villa}, {Zonca}, {Banday}, {Barreiro}, {Bartlett}, {Bartolo},
  {Bedini}, {Bennett}, {Binko}, {Borrill}, {Bouchet}, {Bremer}, {Cabella},
  {Cappellini}, {Chen}, {Colombo}, {Cruz}, {Curto}, {Danese}, {Davies},
  {Davis}, {de Gasperis}, {de Rosa}, {de Troia}, {Dickinson}, {Diego},
  {Donzelli}, {D{\"o}rl}, {Efstathiou}, {En{\ss}lin}, {Eriksen}, {Falvella},
  {Finelli}, {Franceschi}, {Gaier}, {Gasparo}, {G{\'e}nova-Santos}, {Giardino},
  {G{\'o}mez}, {Gruppuso}, {Hansen}, {Hell}, {Herranz}, {Hovest}, {Huynh},
  {Jewell}, {Juvela}, {Kisner}, {Knox}, {L{\"a}hteenm{\"a}ki}, {Lamarre},
  {Leonardi}, {Le{\'o}n-Tavares}, {Lilje}, {Lubin}, {Maggio}, {Marinucci},
  {Mart{\'{\i}}nez-Gonz{\'a}lez}, {Massardi}, {Matarrese}, {Meharga},
  {Melchiorri}, {Migliaccio}, {Mitra}, {Moss}, {N{\o}rgaard-Nielsen}, {Pagano},
  {Paladini}, {Paoletti}, {Partridge}, {Pearson}, {Pettorino}, {Pietrobon},
  {Pr{\'e}zeau}, {Procopio}, {Puget}, {Quercellini}, {Rachen}, {Rebolo},
  {Robbers}, {Rocha}, {Rubi{\~n}o-Mart{\'{\i}}n}, {Salerno}, {Savelainen},
  {Scott}, {Seiffert}, {Silk}, {Smoot}, {Sternberg}, {Stivoli}, {Stompor},
  {Tofani}, {Toffolatti}, {Tuovinen}, {T{\"u}rler}, {Umana}, {Vielva},
  {Vittorio}, {Vuerli}, {Wade}, {Watson}, {White}, \& {Wilkinson}}]{LFI2011}
{Zacchei}, A., {Maino}, D., {Baccigalupi}, C., {et~al.} 2011, A\&A, 536, A5

\end{thebibliography}
\end{document}